\documentclass[a4paper,11pt]{article}
\usepackage{pos}
\usepackage{multirow}
\usepackage{float}
\usepackage{arydshln}
\usepackage{colortbl}
\usepackage{amsmath}
\usepackage{dsfont} 
\usepackage{hyperref}
\usepackage{graphicx}
\usepackage{enumitem}
\usepackage{arydshln}
\usepackage{mathtools}
\usepackage{bbold}
\usepackage{soul}
\usepackage{slashed}
\usepackage{amsmath,bm}
\usepackage{yfonts}
\usepackage{lipsum}
\usepackage{slashed}
\usepackage{mathtools}
\usepackage{physics}
\usepackage{mciteplus}
\usepackage[numbers]{natbib}

\newcommand{\be} {\begin{equation}}
\newcommand{\ee} {\end{equation}}
\newcommand{\bea} {\begin{eqnarray}}
\newcommand{\eea} {\end{eqnarray}}

\newcommand{\cO}{{\mathcal O}}
\newcommand{\cR}{{\mathcal R}}

\newcommand{\cL}{{\mathcal L}}

\newcommand{\cB}{{\mathcal B}} 
 
\newcommand{\llpair}{\bar\ell\ell}
\newcommand{\mmpair}{\bar\mu\mu}

\title{New tests of short-distance dynamics in $b \rightarrow s\bar{\ell} \ell$ decays}

\author*{Arianna Tinari}

\affiliation{Physik-Institut, Universit\"at Zu\"rich, CH-8057 Z\"urich, Switzerland}
\emailAdd{arianna.tinari@physik.uzh.ch}

\abstract{The rare $B \to K^{(*)} \bar{\ell} \ell$ decays exhibit a long-standing tension with Standard Model (SM) predictions, which can be attributed to a lepton-universal short-distance $b \to s \bar{\ell} \ell$ interaction. We present two novel methods to disentangle this effect from long-distance dynamics: one based on the determination of the inclusive $b \to s \bar{\ell} \ell$ rate at high dilepton invariant mass ($q^2\geq 15~{\rm GeV}^2$), the other based on the analysis of the $q^2$ spectrum of the exclusive modes $B \to K^{(*)} \bar{\ell} \ell$ (in the entire $q^2$ range).

Using the first method, we show that the SM prediction for the inclusive $b \to s \bar{\ell} \ell$ rate at high dilepton invariant mass is in good agreement with the result obtained summing the SM predictions for one- and two-body modes ($K$, $K^*$, $K\pi$). This observation allows us to perform a direct comparison of the inclusive $b \to s \bar{\ell} \ell$ rate with data. This comparison shows a significant deficit ($\sim 2\sigma$) in the data, fully compatible with the deficit observed at low-$q^2$ on the exclusive modes. This provides independent evidence of an anomalous $b \to s \bar{\ell} \ell$ short-distance interaction, free from uncertainties on the hadronic form factors. To test the short-distance nature of this effect we use a second method, where we analyze the exclusive $B \to K^{(*)} \bar{\ell} \ell$ data in the entire $q^2$ region. Here, after using a dispersive parametrization of the charmonia resonances, we extract the non-SM contribution to the universal Wilson coefficient $C_9$ for every bin in $q^2$ and for every polarization. The $q^2$- and polarization-independence of the result, and its compatibility with the inclusive determination, provide a consistency check of the short-distance nature of this effect.}

\FullConference{The European Physical Society Conference on High Energy Physics (EPS-HEP2023)\\
 21-25 August 2023\\
Hamburg, Germany\\}

\begin{document}
\maketitle

\section{Introduction}

Exclusive and inclusive $b\to s\llpair$ decays are sensitive probes of physics beyond the Standard Model (SM). 
The flavor-changing neutral-current (FCNC) structure implies a strong suppression of the decay amplitudes within the SM and, correspondingly, enhanced sensitivity to short-distance physics.
On the theory side, the presence of narrow charmonium resonances poses challenges if the invariant mass of the dilepton pair,
$q^2 = (p_{\bar{\ell}} + p_\ell)^2$, is close to the resonance masses.   
This is why precise SM tests are confined to  $q^2 \lesssim 6-8\,$GeV$^2$ (low-$q^2$ region) and $q^2 \gtrsim 14-15\,$GeV$^2$ (high-$q^2$ region).  
On the experimental side, in the last few years measurements of rates and angular distributions of the exclusive $B \to K^{(*)}\mmpair$ decays by LHCb~\cite{LHCb:2013ghj,LHCb:2014cxe} have shown significant tensions with the corresponding SM predictions, especially in the low-$q^2$ region.  

The goal of this study is to better understand the nature of this tension by trying to disentangle long-distance dynamics from possible short-distance dynamics using two different approaches~\cite{Isidori:2023unk, inprep}.

\section{\texorpdfstring{First approach: semi-inclusive $b \rightarrow s \bar{\ell} \ell$ transitions at high $q^2$}{}}
The effective Lagrangian valid below the electroweak scale relevant to $b\to s\llpair$ transitions 
is conventionally written as
\bea
	\cL^{b\to s \llpair}_{\rm eff} &=& \frac{4G_F}{\sqrt{2}}  \frac{\alpha_e}{4\pi}\left(V^*_{ts}V_{tb}\sum_i C_i \cO_i + \text{h.c.} \right)+ 	\cL^{N_f =5}_{\rm QCD\times  QED} \,,
	\label{eq:bsll}
\eea
where we have used CKM unitarity, and neglected the tiny $O(V^*_{us}V_{ub})$ terms. 

The only $\cO_i$ with $b\to s\llpair$ matrix elements that are non-vanishing at tree level are the electric-dipole operator $\cO_7$ and the two FCNC semileptonic operators	$\cO_9$ and $\cO_{10}$:
\be
	\cO_7 =\frac{m_b}{e}(\overline{s}_L\sigma_{\mu\nu} b_R)F^{\mu\nu}\,, \quad 
\cO_{9}=(\overline{s}_L\gamma_{\mu} b_L)(\overline{\ell} \gamma^{\mu}\ell)\,, \quad
\cO_{10}=(\overline{s}_L\gamma_{\mu} b_L)(\overline{\ell} \gamma^{\mu}\gamma_5\ell)\,.		
\ee

We find it convenient to perform a change of basis $\{ \cO_{9},  \cO_{10} \} \to \{ \cO_V,  \cO_L \}$, where 
\be 
	\!\cO_V =(\overline{s}_L\gamma_{\mu} b_L)(\overline{\ell} \gamma^{\mu}\ell)\,, \quad
	\cO_L =(\overline{s}_L\gamma_{\mu} b_L)(\overline{\ell}_L \gamma^{\mu}\ell_L)\,,
 \label{eq:OVOL}
\ee
such that 
$
 C_V = C_9 + C_{10}\,$ and $  C_L = -2 C_{10}\,.
$
The new basis allows us to separate effective interactions that originate from different underlying dynamics, and behave differently in the evolution 
from high scales ($\mu_0 \sim m_t$) down to low scales ($\mu_b \sim m_b$).
Throughout the computation we use the following values:
\be
C_L(m_b)=8.38\pm0.04, \qquad C_V(m_b)=-0.01\pm0.26.
\ee

Our goal is to compare the inclusive rate in the high-$q^2$ region with a semi-inclusive determination based on data, as an inclusive measurement is not yet available. In order to validate this procedure, we first show that, within the SM, the sum of a few exclusive modes (the leading $B^0 \to K^0 \bar{\ell} \ell$, $B^0 \to K^{*0} \bar{\ell} \ell$, and the sub-leading $B^0 \to K \pi \bar{\ell} \ell$) can approximate well the inclusive rate in the high-$q^2$ region. 
As pointed out in~\cite{Ligeti:2007sn}, a convenient way to estimate the inclusive rate is by computing the ratio of the FCNC transition and the
$b\to u$ charged-current decay,
\be
 R^{(\ell)}_{\rm incl}(q_0^2) = \displaystyle\int_{q^2_0}^{m_B^2} d q^2 \frac{d \Gamma(B \to X_s \llpair)}{d q^2} \Bigg/
 \displaystyle\int_{q^2_0}^{m_B^2} d q^2 \frac{d \Gamma(B \to X_u \bar\ell\nu)}{d q^2} \,,
 \label{eq:R0}
\ee
where $q_0^2$ is the lower cut on $q^2$ (we choose $q_0^2=15~\rm GeV ^2 $). 
The hadronic structure of the two transitions is very similar, leading to a significant cancellation of non-perturbative uncertainties. For the inclusive rate
$B \to X_u \bar\ell \nu$ we use the measurements from Belle~\cite{Belle:2021ymg}.

To provide an updated numerical prediction of $ R_{\rm incl}(q_0^2)$ within the SM 
we re-express the result of \cite{Ligeti:2007sn} in the $C_{L,V}$ basis:
\be
 R_{\rm incl}(q_0^2) = \frac{ |V_{tb} V^*_{ts}|^2 }{ |V_{ub}|^2 } \left[\cR_L + \Delta \cR_{[q_0^2]} \right]  = \frac{ |V_{tb} V^*_{ts}|^2 }{ |V_{ub}|^2 } \left[ \frac{\alpha_e^2 C_L^2 }{16 \pi^2} + \Delta \cR_{[q_0^2]} \right],
 \label{eq:RinclCVCL}
\ee
where $\Delta \cR_{[q_0^2]}$ is the correction to the limit of purely left-handed interactions dominating $b \to s \bar{\ell} \ell$ and identical hadronic distributions in $b \to u \ell \bar{\nu}$ compared to $b \to s \llpair$. We find:
\be
\cR_L^{\rm SM} = (2.538 \pm 0.024)\times 10^{-5}\,,  \quad 
\Delta \cR^{\rm SM}_{[15]}  = (-0.03 \pm 0.22)\times 10^{-5}\,,
\ee
and with this, we finally obtain for the branching ratio:
\be
\cB( B \to X_s \llpair)_{[15]}^{\rm SM}  =  (4.5 \pm 1.0) \times  10^{-7} = 4.5 \times 10^{-7} \left[ 1 \pm  0.16_{\rm exp} \pm 0.11_{\rm CKM} \pm 0.09_{\rm \Delta \cR}  \right].\qquad
\label{eq:inclusive}
\ee
The branching fractions for the leading exclusive modes in the high-$q^2$ region can be computed using the form factors calculated in Refs.~\cite{Parrott:2022rgu, Horgan:2015vla}. Integrating for $q^2\ge 15$ GeV, we find
\begin{equation}
        \mathcal{B}(B\to K\llpair)^{\rm SM}_{[15]} = \big(1.31\pm 0.12\big)\times 10^{-7}\,, \quad
        \mathcal{B}(B\to K^*\llpair)^{\rm SM}_{[15]} = \big(3.19\pm 0.30\big)\times 10^{-7}\,,
        \label{eq:leadingmodes}
\end{equation}
The subleading $B \to K \pi$ branching ratio is estimated via heavy-hadron chiral perturbation theory. In order to avoid double-counting the resonant contributions from $B\to (K^* \to K \pi) \bar{\ell} \ell$, we only include the $s$-wave contribution assuming $K^*$-dominance for the $p-$wave. We find
\begin{equation}
    \mathcal{B}\big(B\to (K\pi)_s\llpair)^{\rm SM}_{[15]} = (5.8\pm2.5)\times 10^{-8}\,,
    \label{eq:kpi}
\end{equation}
where the narrow-width approximation is used.

Combining \eqref{eq:leadingmodes} and (\ref{eq:kpi}), we 
arrive at the following SM estimate of the semi-inclusive 
branching fraction:
\begin{equation}\label{eq:BRsemincl}
    \sum_i\mathcal{B}(B\to X^i_s\llpair)^{\rm SM}_{[15]} = \big(5.07\pm 0.42\big)\times 10^{-7}\,.
\end{equation}
This result is well-compatible with the truly inclusive estimate 
presented in Eq.~(\ref{eq:inclusive}).

\begin{figure}[ht!]
\centering
\begin{minipage}[t]{.35\textwidth}
    \centering
    \includegraphics[width=0.91\linewidth]{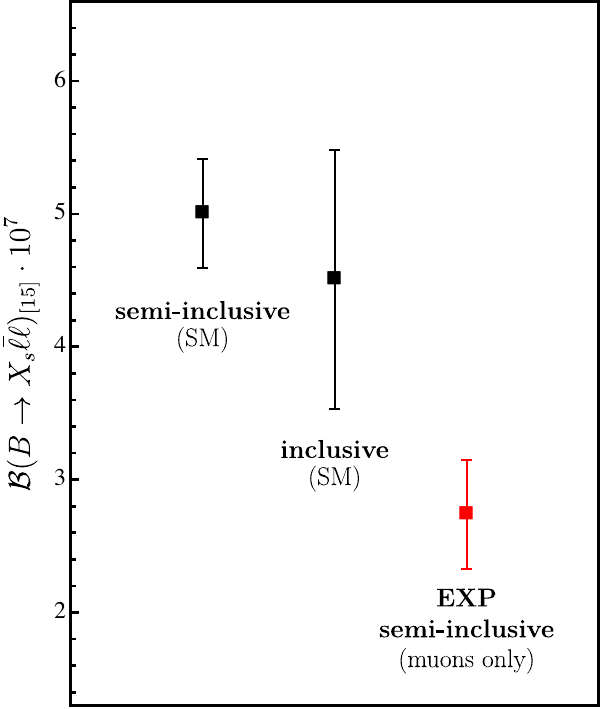}
    \caption{SM predictions vs.~experimental data  for the inclusive 
    branching ratio,  $\cB(B \to X_s \llpair)$, in the region $q^2 \geq 15~{\rm GeV}^2$.
    }
    \label{fig:BR}
\end{minipage}%
\hfil
\begin{minipage}[t]{.55\textwidth}
    \centering
    \includegraphics[width=0.7\linewidth]{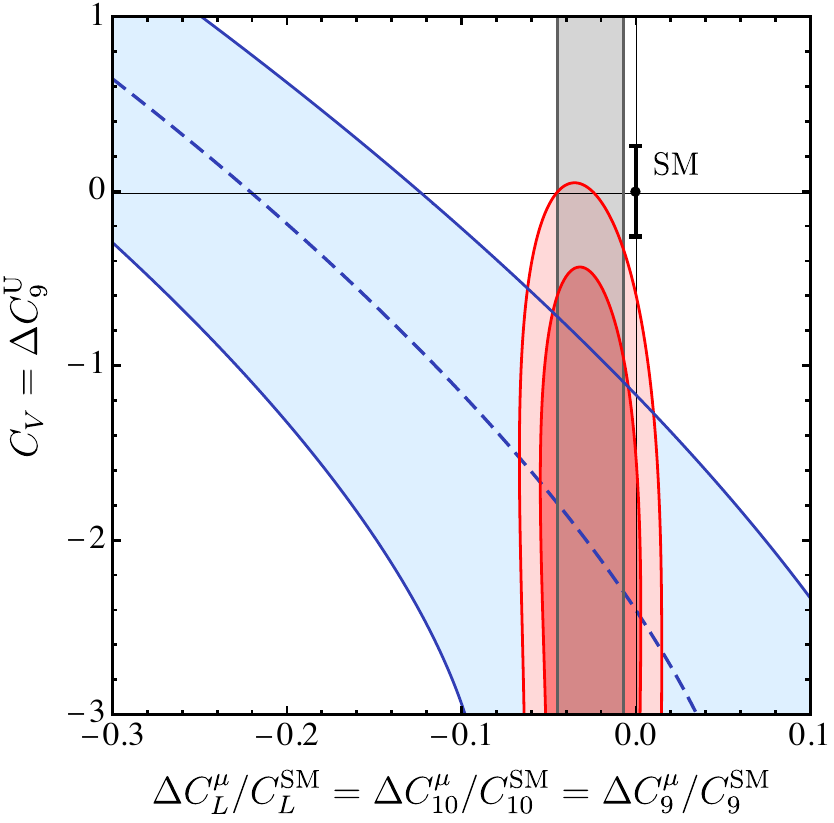}
    \caption{Regions for the Wilson coefficients favored by experimental data. The dark and light red regions give the combined compatibility from the inclusive rate $b \to s \bar{\ell} \ell$ (blue 1$\sigma$-band), LFU tests and $B_s \to \mmpair$ (gray band) at 68\% and 90\% confidence level, respectively.}
    \label{fig:CVCL}
\end{minipage}
\end{figure}

Having established the validity of this procedure, we now compare the inclusive determination (\ref{eq:inclusive}) with a semi-inclusive sum based on the available experimental data from LHCb~\cite{LHCb:2014cxe} (for $\ell= \mu$):
\begin{equation}
        \mathcal{B}(B\to K\mmpair)^{\rm exp}_{[15]}=(8.47\pm 0.50)\times 10^{-8}\,, \quad
        \mathcal{B}(B\to K^*\mmpair)^{\rm exp}_{[15]}=(1.58\pm 0.35)\times 10^{-7}\,.
\end{equation}
Applying a correction factor coming from the sub-leading $B \to K \pi$ mode in (\ref{eq:kpi}), 
we determine the following result:
\begin{equation}
    \sum_i\mathcal{B}(B\to X^i_s\mmpair)^{\rm exp}_{[15]}  = (2.74\pm 0.41) \times 10^{-7}\,.
    \label{eq:BRexp}
\end{equation}
As summarized in Fig.~\ref{fig:BR}, this result is significantly below the 
(consistent) SM predictions in  Eqs.~(\ref{eq:inclusive})
and~(\ref{eq:BRsemincl}). This provides an independent verification of the known suppression in the $b\to s \bar{\mu} \mu$, since, being based on the inclusive rate, it is insensitive to hadronic form factors, and has a different sensitivity to non-perturbative effects associated with charm-rescattering.

 In Fig.~\ref{fig:CVCL} we plot the region in the $C_V$--$C_L$ plane favored by present data (blue dashed line with band). Perturbative and non-perturbative contributions due to charm-rescattering can be accounted for via an effective modification to $C_V$. Assuming $C_L = C_L^{\mathrm{SM}}$, the modification needed is very large; in fact, it is larger than the perturbative estimate of charm-rescattering contributions, and beyond any realistic estimate of non-perturbative charm-rescattering in the high-$q^2$ region, far from the narrow charmonium resonances. 
 If, instead, we allow for a (lepton-flavor non-universal) modification of $C_L$, which can occur only beyond the SM, the discrepancy with the data is more easily explained with a small (naturally lepton-flavor universal) modification to $C_V$. Combining the constraints on $C_L$ coming from LFU tests \citep{LHCb:2022qnv} and $B_s \to \mmpair$ \cite{CMS:2022mgd} (gray band), leads to a preferred region (red region) in the $C_V - \Delta C_L^\mu$ plane that does not include the SM point at the 90\% confidence level.

\section{\texorpdfstring{Second approach: bin-by-bin extraction of $C_9$}{}}
As a further hint to the short-distance nature of this effect, we look at the exclusive modes $B^+ \to K^+ \bar{\mu} \mu$ and $B^0 \to K^{*0} \bar{\mu} \mu$, and show that the Wilson coefficient $C_9$ extracted from data does not show a significant $q^2-$ or helicity-dependence.

In order to compute the theory predictions, we effectively modify $C_9$ by taking into account the perturbative corrections coming from the four-quark operators $\mathcal{O}_{1-6}$. Since in this analysis the entire $q^2$ region is considered, a parametrization of the $c \bar{c}$ resonances is also needed: we use a dispersive approach (see \cite{Khodjamirian:2012rm}), where each resonance ($J/\psi$,  $\psi(2s)$,  $\psi(3770)$, $\psi(4040)$, $\psi(4160)$, and $\psi(4450)$ for the $K$ case, and only $J/\psi$,  $\psi(2s)$ for $K^*$ due to a lack of data) is parameterized by two unknown parameters that are extracted from data. 

In the case of $B \to K \bar{\mu} \mu$, we perform a fit of $C_9$ bin by bin in $q^2$ by using the measured branching ratio by LHCb \cite{LHCb:2014cxe} and more recently by CMS \cite{CMS-PAS-BPH-22-005}, and the form factors computed in \cite{Parrott:2022rgu} for the theory prediction. In the low-$q^2$ region, since the experimental bins used by LHCb and by CMS are the same, we combine the two measurements, whereas in the high-$q^2$ region we carry out two independent fits, using LHCb and CMS data separately. 
In the case of $B \to K^* \bar{\mu} \mu$, we perform the fit from the branching ratio and the angular observables measured by LHCb \cite{LHCb:2016ykl}, using the form factors computed in \cite{Bharucha:2015bzk} for the theory predictions.

\begin{figure}[h!]
    \centering
    \includegraphics[scale=0.37]{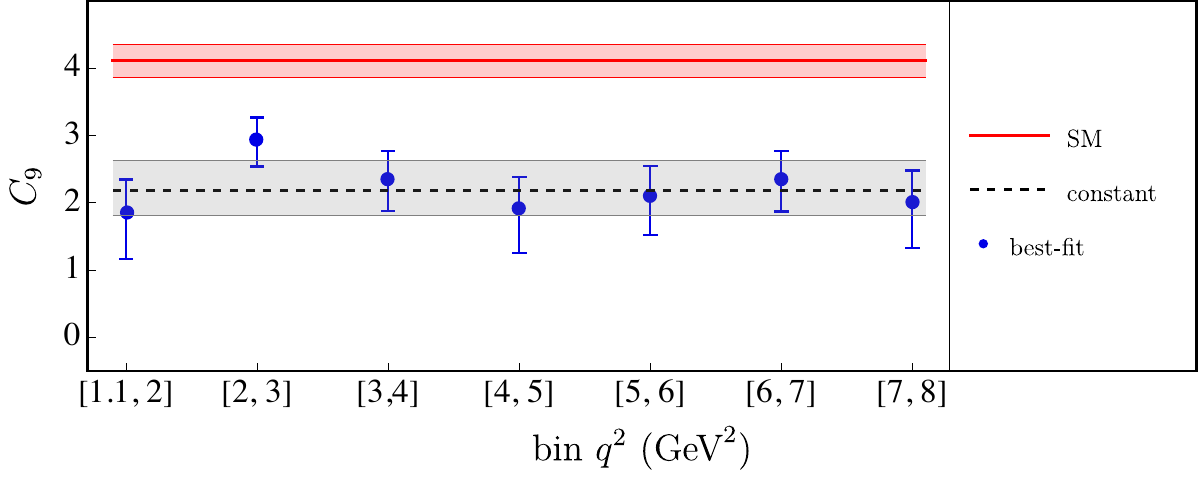}
    \includegraphics[scale=0.37]{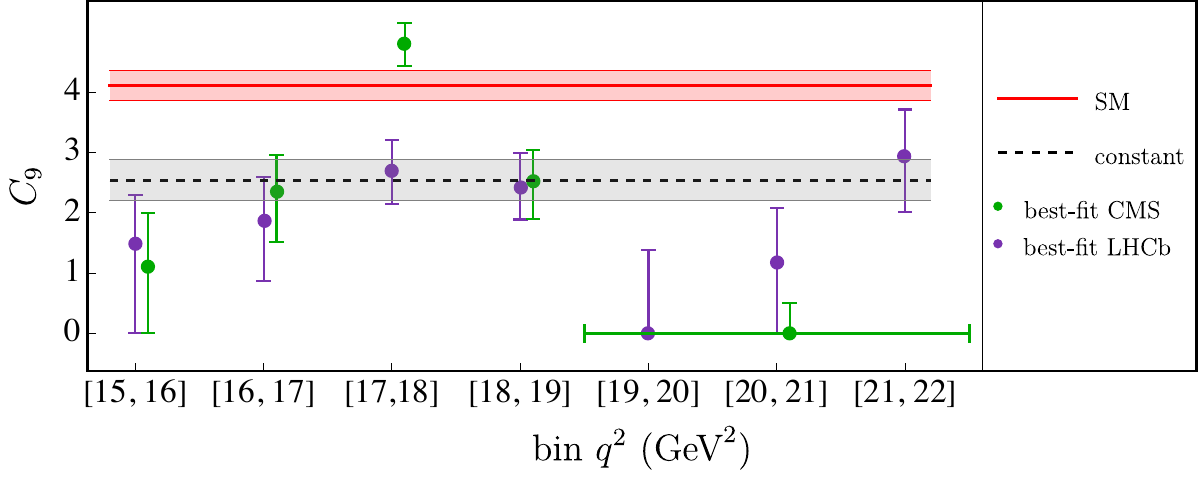}    

    \centering
    \includegraphics[scale=0.37]{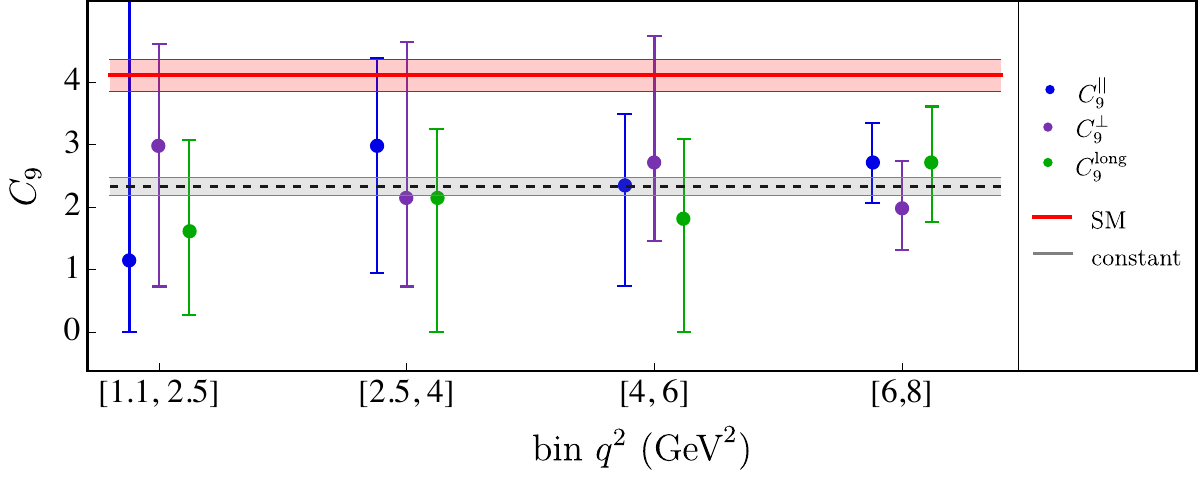}
    \includegraphics[scale=0.37]{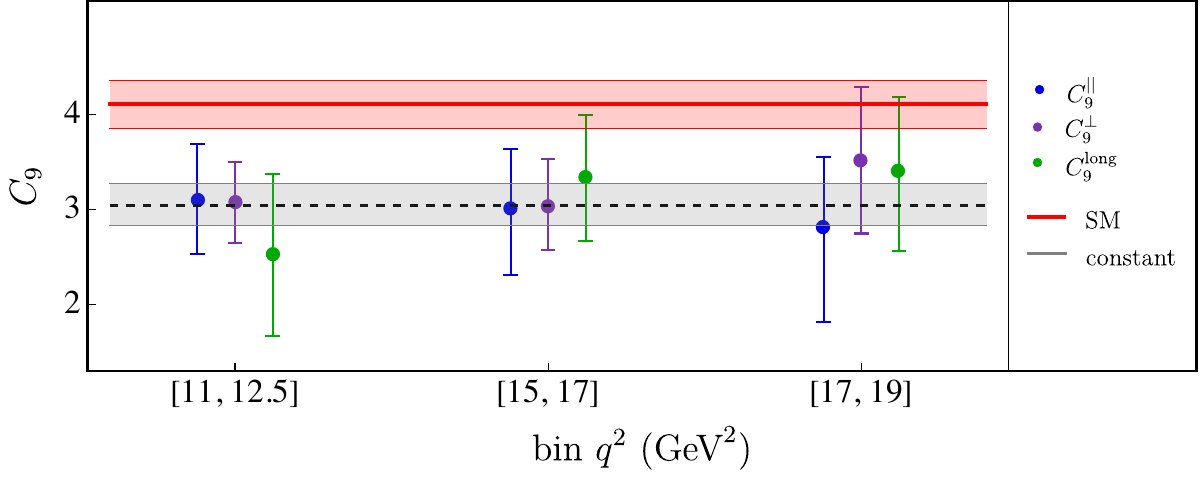}
    \caption{Determinations of $C_9$ in different $q^2$ bins  
    from $B \rightarrow K \mmpair$ (top) and $B \rightarrow K^* \mmpair$ (bottom) data. 
    The red and gray bands denote the SM value and the value extracted assuming 
    a constant ($q^2$-independent) $C_9$, respectively. }
        \label{fig:fit}
\end{figure}

\begin{figure}[h!]
\centering
\begin{minipage}[b]{.45\textwidth}
The results of the fit are shown in Fig.~\ref{fig:fit}. The best-fit results under the assumption of a $q^2$-independent $C_9$ are also shown (gray lines). We do not notice a significant $q^2$- or polarization-dependence, which would be present if we were missing the contribution of dominant long-distance QCD effects.
In Fig.~\ref{fig:final} we show the best-fit results with the assumptions of $q^2$-independent $C_9$, in the low- and high-$q^2$ regions and for the different modes and polarizations. 
\end{minipage}%
\hfil
\begin{minipage}[b]{.5\textwidth}
    \centering
    \includegraphics[scale=0.4]{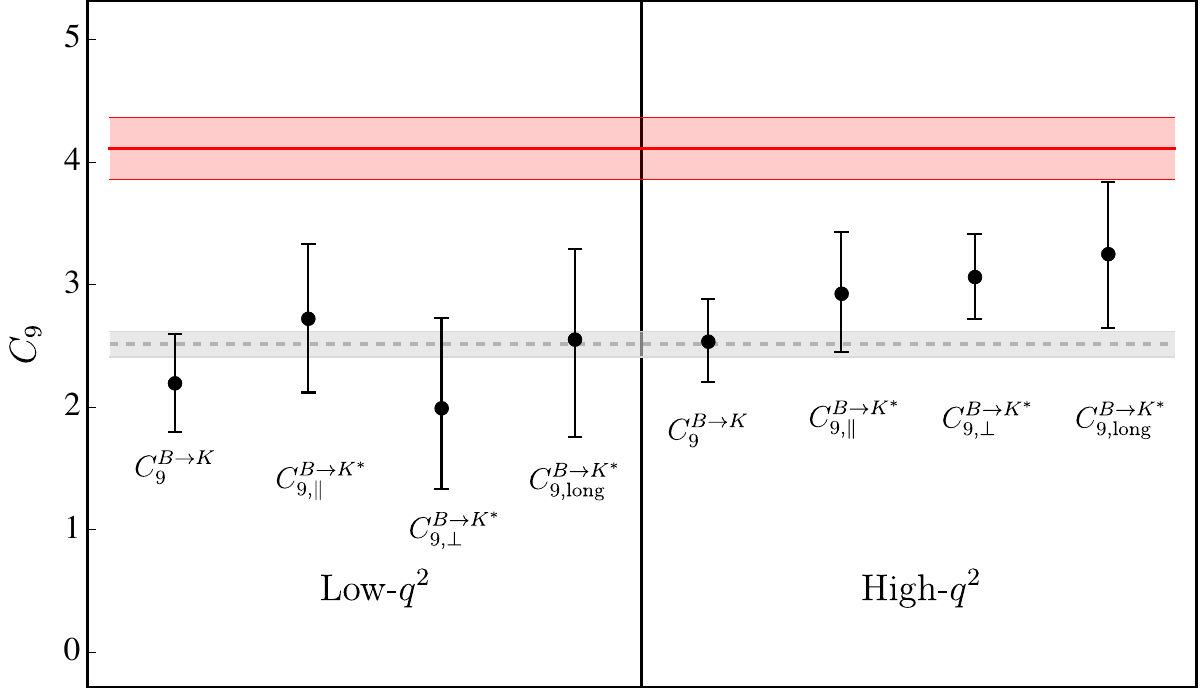}
    \caption{Independent determinations of $C_9$.}
        \label{fig:final}
\end{minipage}
\end{figure}
We also combine $B\to K$ and $B \to K^*$ in the full $q^2$ spectrum (gray dashed line).
These eight independent determinations of $C_9$ are in good agreement with each other, providing another consistency check of the short-distance nature of the tension between the SM and $b \to s \llpair$ data.

\end{document}